\documentclass[12pt]{article}
\usepackage{subfloat}
\pdfoutput = 1
\usepackage{graphics}
\usepackage{graphicx} 
\usepackage{amsmath}
\usepackage{amsfonts}   
\usepackage{amssymb}
\usepackage{multirow}
\usepackage{hhline}
\usepackage{color}
\begin{document}
\date{Today}
\title{{\bf{\Large Holographic insulator/superconductor phase transition by matching method and thermodynamic geometry  }}}
\author{ {\bf {\normalsize Diganta Parai}$^{a}$
\thanks{digantaparai007@gmail.com}},\,
{\bf {\normalsize Debabrata Ghorai}$^{b}
$\thanks{debanuphy123@gmail.com, debabrataghorai@bose.res.in}},\,
{\bf {\normalsize Sunandan Gangopadhyay}$^{b}$
\thanks{sunandan.gangopadhyay@gmail.com, sunandan.gangopadhyay@bose.res.in}}\\
$^{a}$ {\normalsize Indian Institute of Science Education and Research Kolkata}\\{\normalsize Mohanpur, Nadia 741246, India}\\[0.2cm] 
$^{b}$ {\normalsize Department of Theoretical Sciences } \\{\normalsize S.N. Bose National Centre for Basic Sciences }\\{\normalsize JD Block, 
Sector III, Salt Lake, Kolkata 700106, India}\\[0.2cm]
}
\date{}

\maketitle
\begin{abstract}
\noindent In this work, we have analytically analyzed the insulator/superconductor phase transition in the presence of a 5-dimensional $AdS$ soliton background using matching method and thermodynamic geometry approach . We have first employed the matching method to obtain the critical chemical potential. We then move on to investigate the free energy and thermodynamic geometry of this model in 3+1 dimensions. This investigation of the thermodynamic geometry leads to the critical chemical potential of the system from the condition of the divergence of the scalar curvature. We have then compared the value of the critical chemical potential $\mu_{c}$ in dimension $d=5$ obtained from these two different methods, namely, the matching method and the thermodynamic geometry procedure. We have also obtained an expression for the condensation operator using the matching method. Our findings agree very well with the numerical findings in the literature.       
\end{abstract}
\vskip 1cm
\section{Introduction}
\noindent It is very difficult to study strongly coupled systems using the standard techniques of  perturbation theory in condenced matter physics. The $AdS/CFT$ correspondence \cite{adscft1}-\cite{adscft4} allows us to analyze such systems. The duality claims that a 4-dimensional strongly coupled gauge theory is related to a 5-dimensional weakly coupled gravity theory. This fascinating development is very important for theoretical physics, which allows one to map the strongly coupled systems to weakly coupled systems. The weakly coupled system can be tackled by perturbation theory and then using this correspondence, one can get a picture of some of the properties of the strongly coupled  system. 

\noindent The asymptotically $AdS$ black hole spacetime in the bulk can become unstable leading to the condensation of scalar hair below a certain critical temperature. This instability corresponds to a second order phase transition from normal to superconducting state thereby giving birth to the model of the holographic s-wave  superconductor  and  owes  its  origin  to  the  breaking  of  a  local $U(1)$ symmetry near the event horizon of the black hole. A number of investigations have been made in this direction in order to understand various properties of holographic superconductor/metal phase transition in the framework of usual Maxwell electromagnetic theory \cite{ssg}-\cite{horowitz} as well as in Born-Infeld electrodynamics \cite{hs19}-\cite{dg6} which is a non-linear theory of electrodynamics. It has also been realized that there can be a holographic superconductor model in the bulk $AdS$ soliton background which has the ability to describe an insulator/superconductor phase transition. In this case the AdS soliton background becomes unstable to form condensates of the scalar field which is then interpreted as a superconducting phase for the chemical potential  $\mu>\mu_{c}, \mu_{c}$ being the critical chemical potential.

\noindent Another interesting development that has taken place recently is the association of a geometrical structure with thermodynamic systems in equilibrium.  This was first realized through the works in \cite{FW1}-\cite{GP}.  It was shown that one can get a Riemannian metric with an Euclidean signature from the equilibrium state of a thermodynamic system.  The Riemannian scalar curvature can then be computed and captures the details of interactions of the thermodynamic system.  It turns out that this framework based on a geometrical structure gives a handle to study critical phenomena \cite{GP}.  

\noindent In this present work, we have investigated analytically  a holographic model of insulator to superconductor phase transition in $AdS_{5}$ soliton background using the matching method and the formalism of the thermodynamic geometry. We employ the matching method to obtain the behaviour of the matter fields near the tip of the soliton. This in turn is used to compute the critical chemical potential. We obtain the critical chemical potential for a value of the matching point where the near tip and boundary values of the fields are matched. The analysis is based on the probe limit approximation  which  neglects the back reaction of the matter fields on the background space-time geometry, and is carried out for a particular boundary condition. We also compute the condensation operator using the matching method. We  then  proceed  to  compute  the  free  energy  of  this  $3 + 1$-dimensional holographic superconductor. The trick here is to relate the free energy of the theory on the boundary to the value of the on-shell action of the Abelian-Higgs sector of the full Euclidean action with proper boundary terms \cite{CP},\cite{sppg}.  From this, we compute the thermodynamic metric using the formalism of \cite{GP}. The scalar curvature is computed next and the chemical potential at which the scalar curvature diverges is said to be the critical chemical potential in this approach. This chemical potential is then compared with that obtained from the matching method. 

\noindent This paper is organized as follows. In section 2, we discuss the basic set up for the insulator/superconductor phase transition in $AdS_{5}$ soliton background. In section 3, using the probe limit approximation, we compute the critical chemical potential using the matching method,where the matching has been carried out at a point between boundary and the tip of the soliton. In section 4, we analytically obtain the free energy expression in terms of the chemical potential and the charge density. In section 5, we calculate the thermodynamic metric and the scalar curvature. We conclude finally in section 6. \\

\section{Basic set up}
\noindent The model of a holographic insulator/superconductor phase transition with the Einstein-Maxwell-scalar action in five dimensional spacetime can be constructed by considering the following action
\begin{equation}
S=\int d^{5}x \sqrt{-g}\Big[R+\frac{12}{L^2} -\frac{1}{4} F^{\mu \nu} F_{\mu \nu} - (D_{\mu}\psi)^{*} D^{\mu}\psi-m^2 \psi^{*}\psi\Big]
\label{1} 
\end{equation}
where $F_{\mu \nu}=\partial_{\mu}A_{\nu}-\partial_{\nu}A_{\mu}$ is the field strength tensor, $D_{\mu}\psi=\partial_{\mu}\psi-iqA_{\mu}\psi$ is the covariant derivative,  $A_{\mu}$ and $ \psi $ represent the gauge and the scalar fields, $L$ is the radius of $AdS$ spacetime.\\
\noindent When the Maxwell field and scalar field are absent, the above action admits $AdS$ soliton solution \cite{metric1}
\begin{equation}
ds^2=\frac{L^2}{f(r)}dr^2+r^2(-dt^2+ dx^2+ dy^2)+f(r)d{\chi}^2
\label{2}
\end{equation}
with $f(r)$ being given by
\begin{equation}
f(r)=r^2\left(1-\frac{r_{0}^4}{r^4}\right)
\label{3}
\end{equation}
where $r_{0}$ is the tip of the soliton. For the smoothness at the tip, we need to impose a period $\beta=\frac{\pi L}{r_{0}}$ for the coordinate $\chi$. This gives a dual picture of a three dimensional field theory with a mass gap, which resembles an insulator in condensed matter physics \cite{ccp}.\\
\noindent Making the ansatz $\psi=\psi(r)$ and $A_{t}=\phi(r)$, the equations of motion for the scalar field $\psi$ and gauge field $\phi$ read
\begin{equation}
\partial _{r}^2\psi+\left(\frac{\partial _{r}f}{f}+\frac{3}{r}\right)\partial _{r}\psi+\left(-\frac{m^2}{f}+\frac{q^2 \phi^2}{r^2 f} \right )\psi=0 
\label{4}
\end{equation}
\begin{equation}
\partial _{r}^2\phi+\left(\frac{\partial _{r}f}{f}+\frac{1}{r}\right)\partial _{r}\phi-\frac{2q^2 \psi^2}{ f}\phi=0 ~.
\label{5}
\end{equation}
\noindent By introducing a new coordinate $z=\frac{r_{0}}{r}$, the above equations take the form

\begin{equation}
\psi^{\prime\prime}(z)+\left(\frac{F^{\prime}(z)}{F(z)}-\frac{3}{z} \right )\psi^{\prime}(z)+\left(\frac{q^2\phi^2(z)}{r_{0}^2F(z)}-\frac{m^2}{z^2 F(z)}\right)\psi(z)=0
\label{6}
\end{equation}

\begin{equation}
\phi^{\prime\prime}(z)+\left(\frac{F^{\prime}(z)}{F(z)}-\frac{1}{z} \right )\phi^{\prime}(z)+\frac{2q^2\psi^2(z)}{z^2F(z)}\phi(z)=0.
\label{7}
\end{equation}

\noindent Also under the above transformation of coordinates, the spacetime metric (\ref{3}) becomes 

\begin{equation}
f(r)=r_{0}^2\frac{F(z)}{z^2}
\label{8}
\end{equation}
\noindent where $F(z)=1-z^4$. The rescalings $\psi\rightarrow \frac{\psi}{q}$ and $\phi\rightarrow (r_{0}/q)\phi$ allows one to set $q=1$ and $r_{0}=1$. For the rest of analysis we shall set $L=1$. The asymptotic  behaviour of the fields can be written as

\begin{equation}
\psi_{b}(z)=\psi_{-}z^{\Delta_{-}}+\psi_{+}z^{\Delta_{+}}
\label{9}
\end{equation}

\begin{equation}
\phi_{b}(z)=\mu-\rho z^2
\label{10}
\end{equation}
where
\begin{equation}
\Delta_{\pm}=2\pm\sqrt{4+m^2} ~.
\label{11}
\end{equation}

\noindent From the AdS/CFT dictionary, $\psi_{\pm}$ can be interpreted as the expectation value of the condensation operator $\mathcal{O}_{\pm}$ of the dual field theory living in the boundary. We are focusing on the boundary condition in which $\psi_{-}=0$ and $\psi_{+} \neq 0$. In principle, one can do the same analysis with opposite boundary condition that is $\psi_{+}=0$ and $\psi_{-} \neq 0$. This type of boundary condition is required because we want to turn on the condensate without being externally sourced.

\section{The critical chemical potential $\mu_{c}$ from matching method}
\noindent It has been shown numerically that when the chemical potential $\mu$ exceeds a critical value $\mu_{c}$, the condensations of the operators occur. This can be viewed as a superconductor phase. For $\mu<\mu_{c}$ , the scalar field is zero and this can be viewed as an insulator phase \cite{ccp}. Therefore, the critical chemical potential $\mu_{c}$ is the turning point of this holographic insulator/superconductor phase transition. So near $\mu_{c}$, eq.(\ref{7}) reduces to 
\begin{equation}
\phi^{\prime\prime}(z)+\left(\frac{F^{\prime}(z)}{F(z)}-\frac{1}{z} \right )\phi^{\prime}(z)=0.
\label{12}
\end{equation}

\noindent Multiplying by the integrating factor $\frac{F(z)}{z}$, the above equation can be recast as
\begin{eqnarray}
\frac{\mathrm{d} }{\mathrm{d} z}\left \{ \frac{F(z)}{z}\phi^{\prime}(z) \right \}&=&0\nonumber\\
\Rightarrow \frac{F(z)}{z}\phi^{\prime}(z)&=&constant ~.
\label{13}
\end{eqnarray}

\noindent Using the fact that $F(z=1)=0$, we find that the above $constant=0$. Hence the gauge field equation finally reduces to
\begin{eqnarray}
\phi^{\prime}(z)&=&0\nonumber\\
\Rightarrow \phi(z)&=&C=\mu
\label{14}
\end{eqnarray}
where the constant of integration $C$ gets fixed from the asymptotic behaviour of the field $\phi(z)$(eq.(\ref{10})).\\
\noindent The Taylor series expansions of the fields near the tip read
\begin{equation}
\psi _{t}(z)=\psi(1)+\psi^{\prime}(1)(z-1)+\frac{\psi^{\prime\prime}(1)}{2}(z-1)^2+...
\label{15}
\end{equation}
\begin{equation}
\phi _{t}(z)=\phi(1)+\phi^{\prime}(1)(z-1)+\frac{\phi^{\prime\prime}(1)}{2}(z-1)^2+... ~~.
\label{16}
\end{equation}

\noindent We shall now determine the undetermined coefficients in eq.(s)(\ref{15}, \ref{16}) using eq.(s)(\ref{6}, \ref{7}). These read
\begin{eqnarray}
\psi^{\prime}(1)&=&\frac{\psi(1)}{4}\left \{  \phi^2(1)-m^2 \right \}
\label{17}\\
\psi^{\prime\prime}(1)&=&\psi(1)\left [ \frac{\phi^2(1)}{4}\left \{ 1-\frac{\psi^2(1)}{2} \right \} -\frac{1}{32}\left \{ \phi^2(1)-m^2 \right \}\left \{ 8+m^2-\phi^2(1) \right \}\right ]
\label{18}\\
\phi^{\prime}(1)&=&-\frac{1}{2}\psi^2(1)\phi(1)
\label{19}\\
\phi^{\prime\prime}(1)&=&\frac{\psi^2(1)\phi(1)}{8}\left \{ 8+m^2+\psi^2(1)-\phi^2(1) \right \} ~.
\label{20}
\end{eqnarray}

\noindent Hence the near tip expansions of these fields up to $\mathcal{O}(z-1)^2$ read
\begin{eqnarray}
\psi_{t}(z)&=&\psi(1)\Bigg [ 1+\frac{z-1}{4}\left \{ \phi^2(1)-m^2 \right \}\nonumber\\
&+&\frac{(z-1)^2}{2}\left [ \frac{\phi^2(1)}{4}\left \{ 1-\frac{\psi^2(1)}{2} \right \} -\frac{1}{32}\left \{ \phi^2(1)-m^2 \right \}\left \{ 8+m^2-\phi^2(1) \right \}\right ] \Bigg ]
\label{21}
\end{eqnarray}

\begin{equation}
\phi_{t}(z)=\phi(1)\left [ 1-\frac{(z-1)}{2}\psi^2(1)+\frac{(z-1)^2}{16}\psi^2 (1)\left \{ 8+m^2+\psi^2(1)-\phi^2(1) \right \}\right ] ~.
\label{22}
\end{equation}

\noindent We now proceed to match the near tip expansions of the fields with the asymptotic solution of these fields at any arbitrary point between the tip and the boundary, say $z=\frac{1}{\lambda}$, where $\lambda$ lies between $\left [ 1,\infty  \right ]$.

\noindent The matching conditions are
\begin{equation}
\psi_{t}\left(\frac{1}{\lambda}\right)=\psi_{b}\left(\frac{1}{\lambda}\right),~~\psi_{t}^{\prime}\left(\frac{1}{\lambda}\right)=\psi_{b}^{\prime}\left(\frac{1}{\lambda}\right)
\label{23}
\end{equation}
\begin{equation}
\phi_{t}\left(\frac{1}{\lambda}\right)=\phi_{b}\left(\frac{1}{\lambda}\right),~~\phi_{t}^{\prime}\left(\frac{1}{\lambda}\right)=\phi_{b}^{\prime}\left(\frac{1}{\lambda}\right) ~.
\label{24}
\end{equation}

\noindent From eq.(\ref{24}), we obtain the following relations

\begin{equation}
\psi^4(1)+\psi^2(1)\left \{ 4\left(\frac{4\lambda -3}{\lambda -1}\right)+m^2-\phi^2(1) \right \}=0
\label{25}
\end{equation}
\begin{equation}
\rho =\frac{\lambda \phi(1)\psi^2(1)}{4}\left [ 1+\frac{\lambda -1}{4\lambda }\left \{ 8+m^2+\psi^2(1)-\phi^2(1) \right \} \right ].
\label{26}
\end{equation}

\noindent Near the critical chemical potential, eq.(s)(\ref{25}, \ref{26}) reduce to
\begin{equation}
\psi^2(1)=\left \{ \mu^2-m^2-4\left(\frac{4\lambda -3}{\lambda -1}\right) \right \}
\label{27}
\end{equation}

\begin{equation}
\rho =\frac{(\lambda-1)\mu}{4}\left\{\frac{4(4\lambda-3)}{\lambda-1}+m^2-\mu^2\right\} ~.
\label{28}
\end{equation}

\noindent From eq.(\ref{23}), we get
\begin{eqnarray}
-\lambda \Delta +\frac{1+\Delta (\lambda -1)}{4}\left \{ \phi^2(1)-m^2 \right \}-\frac{\lambda -1}{4\lambda }\left \{ 1+\frac{\Delta (\lambda -1)}{2} \right \}\phi^2(1)\left \{ 1-\frac{\psi^2(1)}{2} \right \}\nonumber\\
+\frac{\lambda -1}{32\lambda }\left \{ 1+\frac{\Delta (\lambda -1)}{2} \right \}\left \{ \phi^2(1)-m^2 \right \}\left \{ 8+m^2-\phi^2(1) \right \}=0.
\label{29}
\end{eqnarray}

\noindent Now near the critical chemical potential,  eq.(\ref{29}) gives with the help of eq.(\ref{27})

\begin{eqnarray}
\frac{3(\lambda -1)}{32\lambda }\left \{ 1+\frac{\Delta (\lambda -1)}{2} \right \}\mu^4+\left [ \left \{ \frac{1+\Delta (\lambda -1)}{4} \right \} -\frac{\lambda -1}{16\lambda }\left \{ 1+\frac{\Delta (\lambda -1)}{2} \right \}\left \{ m^2+8\left ( \frac{4\lambda -3}{\lambda -1} \right ) \right \}\right ]\mu^2\nonumber\\
-\left [ \lambda \Delta +m^2\left \{ \frac{1+\Delta (\lambda -1)}{4} \right \}+\frac{\lambda -1}{32\lambda }\left \{ 1+\frac{\Delta (\lambda -1)}{2} \right \} m^2(8+m^2)\right ]=0~. \nonumber \\
\label{30}
\end{eqnarray}
To estimate the critical chemical potential we now need to solve eq.(\ref{30}). For $m^2=0$ we have $\Delta=\Delta_{+}=4$. Setting $\lambda=2$ the above equation reads
\begin{equation}
9\mu^4-160\mu^2-512=0 ~.
\label{30a}
\end{equation}
The solution of the above equation is $\mu=4.533$ which agrees with the numerical value 3.404 \cite{qjb}.

\noindent We now want to derive an expression for the condensation operator.  From eq.(\ref{23}), we obtain
\begin{equation}
\psi _{+}=\frac{\lambda ^3}{4}\psi (1)\left [ \frac{\phi ^2(1)}{4}+\frac{1-\lambda }{\lambda }\left [ \frac{\phi ^2(1)}{4}\left \{ 1-\frac{\psi ^2(1)}{2} \right \}-\frac{1}{32}\phi^2(1)\left \{ 8-\phi^2(1) \right \}  \right ] \right ]~.
\label{37}
\end{equation}
\noindent Using the map $\psi_{+}=\left \langle \mathcal{O}_{+} \right \rangle$ and eq.(\ref{27}), the above expression near the critical chemical potential takes the form
\begin{equation}
\left \langle \mathcal{O}_{+} \right \rangle= \lambda ^2\left(\frac{7\lambda-6}{16} \right )\mu^2\left [1+\frac{\lambda -1}{2(7\lambda -6)}m^2 -\frac{3(\lambda -1)}{8(7\lambda -6)}\mu^2 \right ]\sqrt{\mu^2-\mu_c^2 }
\label{38a}
\end{equation}
where the critical chemical potential $\mu_{c}$ reads 
\begin{equation}
\mu_{c}=\sqrt{\left \{m^2+\frac{4(4\lambda -3)}{\lambda -1}  \right \}} ~.
\label{38b}
\end{equation} 
For $m^2=0$ and $\lambda=2$, we find $\mu_c= 4.472$. This is in close agreement with our earlier finding.
In Table \ref{tab2} we present values of $\mu_{c}$ obtained from eq.(s)(\ref{30},\ref{38b}). Near the critical chemical potential, the expression of condensation operator reads using eq.(s)(\ref{38a},\ref{38b})
\begin{eqnarray}
\left \langle \mathcal{O}_{+} \right \rangle &=& \lambda ^2\left(\frac{7\lambda-6}{16} \right )\mu_{c}^2\left [1+\frac{\lambda -1}{2(7\lambda -6)}m^2 -\frac{3(\lambda -1)}{8(7\lambda -6)}\mu_{c}^2 \right ]\sqrt{2\mu_{c}}\sqrt{\mu-\mu_c }\nonumber\\
 &=&\frac{\lambda ^2}{16\sqrt{2}}\left ( 2\lambda -3+\frac{\lambda -1}{4}m^2 \right )\left [ m^2+4\left( \frac{4\lambda -3}{\lambda -1}\right ) \right ]^{\frac{3}{4}}\sqrt{\mu-\mu_{c}} \nonumber \\
&\equiv& \gamma \sqrt{\mu-\mu_{c}} ~.
\label{38c}
\end{eqnarray} 
For $m^2=-\frac{15}{4}$ and $\lambda=2.7$, we get $\left \langle \mathcal{O}_{+} \right \rangle=1.940\sqrt{\mu-\mu_{c}}$ which agrees reasonably well with the Sturm-Liouville analytic result $\left \langle \mathcal{O}_{+} \right \rangle=1.801\sqrt{\mu-\mu_{c}}$  in \cite{rhh}. This result shows that the phase transition between the s-wave holographic insulator and superconductor belongs to second order and the critical exponent of the system takes the mean-field value 1/2. In Table 1, we present the values of the condensation operator $\left \langle \mathcal{O}_{+} \right \rangle$ for $m^2=-\frac{15}{4},~0$ obtained by matching the near tip and asymptotic values of the fields at different points parametrized by $\lambda$.  \\

\begin{table}[h]
\caption{Value of condensation operator $ \langle \mathcal{O}_{+} \rangle = \gamma \sqrt{\mu-\mu_{c}} $ with different values of $\lambda$ }
\centering                          
\begin{tabular}{|c|c|c| }            
\hline
$\lambda$& $\gamma$ for $m^2=-\frac{15}{4},~\Delta_{+}=\frac{5}{2}$ & $\gamma$ for $m^2=0,~\Delta_{+}=4$ \\
\hline
2.6 & 1.574 & 5.863\\
\hline
2.7&1.940&6.856\\
\hline
\end{tabular}
\label{tab1}
\end{table}

\section{Holographic free energy}
\noindent In this section, we shall calculate the free energy at zero temperature of the field theory living on the boundary of the (4+1)-dimensional bulk theory.
\noindent To proceed further, we first write down the action for the Abelian Higgs sector
\begin{equation}
S_M=\int d^{5}x \sqrt{-g}\Big[ -\frac{1}{4} F^{\mu \nu} F_{\mu \nu} - (D_{\mu}\psi)^{*} D^{\mu}\psi-m^2 \psi^{*}\psi\Big] ~.
\label{31} 
\end{equation}

\noindent Using the same ansatz $\psi=\psi(r)$ and $A_{t}=\phi(r)$ and setting $r_{0}=1$ and $q=1$, we get

\begin{equation}
S_{M}=\int  d^{5}x\left [\frac{F(z)\phi^{\prime2}(z)}{2z} -\frac{F(z)\psi^{\prime2}(z)}{z^3}+\frac{\psi^2(z)\phi^2(z)}{z^3}-\frac{m^2\psi^2}{z^5}  \right ].
\label{32}
\end{equation}

\noindent Applying the boundary condition $(F(1)=0)$ and the equations of motion (\ref{6}),(\ref{7}), we obtain the on-shell value of the action $S_{o}$ to be

\begin{equation}
S_{o}=\int d^4x\left [ -\frac{F(z)\phi(z)\phi^{\prime}(z)}{2z} \mid _{z=0}+\frac{F(z)\psi(z)\psi^{\prime}(z)}{z^3}\mid _{z=0} -\int_{0}^{1}\frac{\phi^2(z)\psi^2(z)}{z^3} dz\right ].
\label{33}
\end{equation}
\noindent Setting $m^2=0$ and substituting the asymptotic behavior of  $\phi(z)=\mu-\rho z^2$ and $\psi(z)=\psi_{-}+\psi_{+}z^4$ in the first two diverging terms of above action, we get
\begin{equation}
S_{o}=\int d^4x\left [ \mu \rho +4 \psi_{+}\psi_{-}-\int_{0}^{1}\frac{\phi^2(z)\psi^2(z)}{z^3}dz \right ].
\label{34}
\end{equation}

\noindent The free energy of  the 3+1-dimensional boundary field theory can now be obtained by
\begin{eqnarray}
\Omega&=&-T S_{o}\nonumber\\
&=&\beta T V_{3}\left [ -\mu \rho -4\psi _{+}\psi _{-} +I\right ]\nonumber\\
&=&\beta T V_{3}\left [ -\mu \rho  +I\right ] 
\label{35}
\end{eqnarray} 

\noindent where in the second equality we have set $\int d^4x= \beta V_{3}$ , $V_{3}$ being the volume of the 3-dimensional space of the boundary, and in the last equality we have used the fact that $\psi_{-}=0$. The integral $I$ reads

\begin{eqnarray}
I&=&\int_{0}^{1}\frac{\phi ^{2}(z)\psi ^{2}(z)}{z^3}\nonumber\\
&=&\int_{0}^{\frac{1}{\lambda}}\frac{\phi_{b} ^{2}(z)\psi_{b} ^{2}(z)}{z^3}+\int_{\frac{1}{\lambda}}^{1}\frac{\phi_{t} ^{2}(z)\psi_{t} ^{2}(z)}{z^3} \nonumber \\
&\equiv& I_{1}+I_{2} ~.
\label{36}
\end{eqnarray}
\noindent To evaluate the integral we replace $\psi_{b}$, $\phi_{b}$ from eq.(s)(\ref{9},\ref{10}) and $\psi_t$ ,$\phi_t$ from eq.(s)(\ref{21},\ref{22}). Now $I_{1}$ equals to
\begin{eqnarray}
I_{1}&=&\int_{0}^{\frac{1}{\lambda}}\frac{\phi_{b} ^{2}(z)\psi_{b} ^{2}(z)}{z^3}\nonumber\\
&=&\int_{0}^{\frac{1}{\lambda}}\frac{\left ( \mu -\rho z^2 \right )^{2}\psi _{+}^2 z^{2\Delta }}{z^3}\nonumber\\
&=&\int_{0}^{\frac{1}{\lambda}}\left ( \mu^2 -2\mu \rho z^2+\rho ^2z^4 \right )^{2}\psi _{+}^2 z^{5} ~.
\label{39}
\end{eqnarray}

\noindent The evaluation of integral $I_{2}$ gives
\begin{eqnarray}
I_{2}&=&\int_{\frac{1}{\lambda}}^{1}\frac{\phi_{t} ^{2}(z)\psi_{t} ^{2}(z)}{z^3}dz\nonumber\\
&=&\int_{\frac{1}{\lambda}}^{1}\frac{dz}{z^3}\psi^2(1)\left [ 1+\frac{z-1}{4}\phi^2(1)+\frac{(z-1)^2}{2}\left [ \frac{\phi^2(1)}{4} \left \{ 1-\frac{\psi^2(1)}{2} \right \}-\frac{1}{32}\phi^2(1)\left \{ 8-\phi^2(1) \right \}\right ] \right ]^{2}\nonumber\\
&\times& \phi^2(1)\left [ 1-\frac{z-1}{2}\psi^2(1)+\frac{(z-1)^{2}}{16} \psi^2(1)\left \{ 8+\psi^2(1)-\phi^2(1) \right \}\right ]^{2} ~.
\label{40}
\end{eqnarray}
\noindent Using eq.(\ref{27}) and $\phi(1)=\mu$, we obtain from the above equation 
\begin{eqnarray}
I_{2}=\int_{\frac{1}{\lambda}}^{1}dz\frac{\mu^2}{z^3}\left \{\mu^2-\frac{4(4\lambda -3)}{\lambda -1}\right \}\left [ 1+\frac{z-1}{4}\mu^2+\frac{(z-1)^2}{64}\mu^2\left \{ \frac{16(4\lambda -3)}{\lambda -1}-3\mu^2 \right \} \right ]^{2}\nonumber\\
\left [ 1-\left \{\mu^2-\frac{4(4\lambda -3)}{\lambda -1}\right \}\left \{ \frac{z-1}{2}+\frac{2\lambda -1}{4(\lambda -1)}(z-1)^{2} \right \} \right ]^{2} ~.
\label{41}
\end{eqnarray}
\noindent Substituting eq.(\ref{39}) and eq.(\ref{41}) in eq.(\ref{35}), the analytical expression for the free energy in terms of the chemical potential and charge density reads
\begin{eqnarray}
\frac{\Omega }{V_{3}}&=&-\mu \rho +\int_{0}^{\frac{1}{\lambda}}dz\left ( \mu^2 -2\mu \rho z^2+\rho ^2z^4 \right )^{2} z^{5}\left ( \frac{\lambda ^3}{4} \right )^{2}\left ( \frac{7\lambda -6}{4\lambda } \right )^{2}\left \{\mu^2-\frac{4(4\lambda -3)}{\lambda -1}  \right \}\nonumber \\
&\times& \mu^4\left [ \frac{3(\lambda -1)}{8(7\lambda -6)}\mu^2-1 \right ]^{2}
+\int_{\frac{1}{\lambda}}^{1}dz\frac{\mu^2}{z^3}\left \{\mu^2-\frac{4(4\lambda -3)}{\lambda -1}\right \}\left [1+\frac{z-1}{4}\mu^2+\frac{(z-1)^2}{64}\mu^2 \right. \nonumber \\
&\times& \left. 
\left \{ \frac{16(4\lambda -3)}{\lambda -1}-3\mu^2 \right \} \right ]^{2} 
\left [ 1-\left \{\mu^2-\frac{4(4\lambda -3)}{\lambda -1}\right \}\left \{ \frac{z-1}{2}+\frac{2\lambda -1}{4(\lambda -1)}(z-1)^{2} \right \} \right ]^{2}~. \nonumber \\
\label{42}
\end{eqnarray}
From the above result, the holographic free energy in terms of the chemical potential and charge density reads (for $\lambda=2$) 
 \begin{eqnarray}
 \frac{\Omega }{V_{3}}=\mu^2 (-20 + \mu^2) (0.400365 - 0.0439475 \mu^2 -0.010626 \mu^4 + 0.00120132 \mu^6 \nonumber\\
     +0.000133451 \mu^8 -0.0000161131 \mu^{10} + 3.97339\times 10^{-7} \mu^{12})- \mu \rho\nonumber\\ +0.0000228882 \mu^4(21.3333 -\mu^2)^2 (-20 + \mu^2) (\mu^2 - 0.375 \mu \rho + 0.0375 \rho^2)~.
 \label{42a}
 \end{eqnarray}

\noindent In the next section, we shall make use of these results to investigate the thermodynamic geometry of the system.

\section{Critical chemical potential from thermodynamic geometry analysis}
\noindent With the above results in hand, we now proceed to investigate the thermodynamic geometry of this holographic superconductor. The thermodynamic metric is defined as
\begin{equation}
g_{ij}=-\frac{1}{\mu}\frac{\partial ^2\omega (\mu,\rho)}{\partial x^{i}\partial x^{j}}
\label{43}
\end{equation}
where $\omega=\frac{\Omega}{V_{3}}$, $x^{1}=\mu$ and $x^{2}=\rho$. Note that we have modified the standard definition of thermodynamic metric $g_{ij}=-\frac{1}{T}\frac{\partial ^2\omega (T,\rho)}{\partial x^{i}\partial x^{j}}$ \cite{tg}. Since the chemical potential plays the main role of the phase transition between insulator to superconductor, we have replaced temperature $T$ by the chemical potential $\mu$ in the definition of the thermodynamic metric. 

\noindent The scalar curvature of a general metric 

\begin{equation}
ds_{th}^2= g_{11}\left ( dx_{1} \right )^{2}+2g_{12}dx_{1}dx_{2}+g_{22}\left ( dx_{2} \right )^{2}
\label{44}
\end{equation}

\noindent is given by \cite{GP}
\begin{equation}
R= \frac{-1}{\sqrt{g}}\left [ \frac{\partial }{\partial x^1}\left ( \frac{g_{12}}{g_{11}\sqrt{g}} \frac{\partial g_{11}}{\partial x^2}-\frac{1}{\sqrt{g}}\frac{\partial g_{22}}{\partial x^1}\right ) +\frac{\partial }{\partial x^2}\left ( \frac{2}{\sqrt{g}}\frac{\partial g_{22}}{\partial x^2} -\frac{1}{\sqrt{g}}\frac{\partial g_{11}}{\partial x^2}-\frac{g_{12}}{g_{11}\sqrt{g}}\frac{\partial g_{11} }{\partial x^1}\right )\right ].
\label{45}
\end{equation}

\noindent A singularity in $R$ can be found by checking whether the denominator of the right-hand side of eq.(\ref{45}) vanishes. The condition of the divergence of $R$ is $g=0$ which reads
\begin{equation}
g_{\mu\mu}g_{\rho\rho}-g_{\mu\rho}^2=0 ~.
\label{46}
\end{equation} 

\noindent The chemical potential for which the scalar curvature diverges can be obtained by solving eq.(\ref{46}) and eq.(\ref{28}) simultaneously. \\
This chemical potential is said to be the critical chemical potential. We take the greatest root as the critical chemical potential.  For the case $m^2=0$, $\lambda=2$, the thermodynamic metric components are

\begin{eqnarray}
g_{\mu\mu}=\frac{1}{\mu}(16.0146 + \mu^2 (-15.3518 - 0.0631753 \mu^8 + 0.0043789 \mu^{10} - 0.0000953613 \mu^{12}\nonumber\\- 1.5625 \mu \rho + 0.47168 \mu^3 \rho -0.0387268 \mu^5 \rho + 0.000944138 \mu^7 \rho +0.09375 \rho^2\nonumber\\+ \mu^2 (1.19283 - 0.0336914 \rho^2) + \mu^6 (0.261182 - 0.0000772476 \rho^2)\nonumber\\
+ \mu^4 (0.263449 + 0.00301208 \rho^2)))
\label{46a}
\end{eqnarray}

\begin{equation}
g_{\rho\rho}=-1.71661\times10^{-6} \mu^3 (21.3333 - \mu^2)^2 (-20 + \mu^2)
\label{46b}
\end{equation}

\begin{eqnarray}
g_{\mu\rho}=\frac{1}{\mu} - 0.390625 \mu^3 + 0.0786133 \mu^5 - 
 0.00484085 \mu^7 + 0.0000944138 \mu^9 + 0.0625 \mu^2 \rho\nonumber\\
 - 0.0134766 \mu^4 \rho + 0.000860596 \mu^6 \rho - 0.0000171661 \mu^8 \rho ~.
\label{46c}
\end{eqnarray}
Substituting the metric components in eq.(\ref{46}) and using  eq.(\ref{28}), we obtain
\begin{eqnarray}
1 - 0.406478 \mu^4+ 0.267053 \mu^6 +0.042373 \mu^8 - 0.0378987 \mu^{10} + 0.00256999 \mu^{12}\nonumber\\
+ 0.000992963 \mu^{14} - 0.00020611 \mu^{16} + 0.0000172084 \mu^{18} -7.48337\times10^{-7} \mu^{20}\nonumber\\
 + 1.67712\times10^{-8} \mu^{22}-1.53569\times10^{-10} \mu^{24}=0.
\label{47}
\end{eqnarray}
The solution of the above equation gives the critical chemical potential, which is $\mu_c= 4.703$. We have compared the critical chemical potential obtained from different equations in Table \ref{tab2}. From the Table, we observe that the value of the critical chemical potential agrees quite well from the two analytical methods we have employed in our analysis and also agrees well with the numerical value.

\begin{table}[h]
\caption{Critical chemical potential ($\mu_{c}$) with different values of $\lambda$ for $m^2=0$ and $\Delta=\Delta_{+}=4$ (Numerical value $\mu_{c}=3.404$  \cite{qjb})}
\centering                          
\begin{tabular}{|c|c|c|c| }            
\hline
$\lambda$&\multicolumn{2}{c|}{$\mu_{c}$ from matching method }& $\mu_{c}$ from divergence of $R$\\
\hhline{~--~}
&From the root of eq.(\ref{30})& From expression of $\mu_c$ (eq.(\ref{38b})) & \\

\hline
2&4.533&4.472&4.703\\
\hline
3&4.340&4.242&4.510\\
\hline
4&4.276&4.163&4.454\\
\hline
5&4.244&4.123&4.428\\
\hline
\end{tabular}
\label{tab2}
\end{table}

\section{Conclusions}
\noindent We now summarize our findings in this work. Using the formalism of matching method and thermodynamic geometry, we have investigated analytically a holographic insulator/superconductor phase transition in the $AdS_{5}$ soliton background. The set up we have considered is that of a 5-dimensional $AdS$ soliton background with the matter field coupled with Maxwell electrodynamics. For $\mu<\mu_{c}$, the $AdS$ soliton background is stable and the dual field theory can be interpreted as an insulator where as for $\mu>\mu_{c}$, the $AdS$ soliton background will be unstable to forming condensates of the scalar field which is interpreted as the superconducting phase in the dual field theory. Using this basic idea, we have calculated the critical chemical potential and condensation operator for $m^2=0$ with a particular boundary condition. The near tip expressions obtained by the matching method plays a crucial role in obtaining the free energy of the holographic superconductor. This in turn is used to compute the thermodynamic geometry. It is observed that the results for the critical chemical potential obtained from the two approaches , namely, the matching method and the thermodynamic geometry method agree with each other. We also obtain an expression for the condensation operator using the matching procedure. The analytical results also match very well with the numerical findings in the literature.  

\section*{Acknowledgments}
DP would like to thank CSIR for financial support.

\end{document}